\documentclass[prd,twocolumn]{revtex4}

\usepackage{graphicx}
\usepackage{dcolumn}
\usepackage{bm}
\usepackage{color}
\usepackage{amsmath}
\usepackage{hyperref}
\def\sec#1{\section{#1} }
\def\ssec#1{\subsection{#1} }
\def\sssec#1{\subsubsection{#1} }
\def\R{\\\\} 
\def\({\left(}
\def\){\right)}
\def\[{\left[}
\def\]{\right]}
\def\a{\alpha}
\def\b{\beta}
\def\f#1#2{\frac{#1}{#2}}
\def\g{\gamma}
\def\d{\partial}
\def\de{\delta}

\def\del{\nabla}
\def\ep{\epsilon}

\def\te{\tilde{\epsilon}}
\def\e{\eta}
\def\h#1{\hat{#1}}

\def\l{\lambda}
\def\L{\Lambda}
\def\m{\mu}
\def\n{\nu}

\def\O#1{{\cal O}({#1}) }
\def\p{\pi}
\def\r{\rho}
\def\s{\sigma}
\def\t{\tau}
\def\th{\theta}

\def\ph{\phi}

\def\<{\langle}
\def\>{\rangle}

\providecommand{\abs}[1]{\lvert#1\rvert} 

\definecolor{orange}{rgb}{1,0.5,0}
\definecolor{test}{rgb}{.5,0.5,.5}

\begin{document}
\preprint{preprint no.}

\title{It's Hard to Learn How Gravity and Electromagnetism Couple}

\author{Yi-Zen Chu}
\author{David M. Jacobs}%
 \author{Yifung Ng}%
 \author{Glenn D. Starkman}%
\affiliation{CERCA, Physics Department, Case Western Reserve University, Cleveland, OH 44106-7079\\}%


\begin{abstract}
We construct the most general effective Lagrangian coupling gravity and electromagnetism up to mass dimension 6 by enumerating all possible non-minimal coupling terms respecting both diffeomorphism and gauge invariance. In all, there are only two unique terms after field re-definitions; one is known to arise from loop effects in QED while the other is a parity violating term which may be generated by weak interactions within the standard model of particle physics.  We show that neither the cosmological propagation of light nor, contrary to earlier claims, solar system tests of General Relativity are useful probes of these terms.  These non-minimal couplings of gravity and electromagnetism may remain a mystery for the foreseeable future.
\end{abstract}

\pacs{Valid PACS appear here}
\maketitle

\section{Introduction}\label{sec:I}

The coupled Maxwell-Einstein system has been extensively studied.  Beyond the classical level, non-minimal coupling of the electromagnetic and gravitational interactions may be described by an effective Lagrangian built perturbatively from electromagnetic and geometric tensors.  This is not merely an academic pursuit as such terms are known to be generated within quantum electrodynamics (QED) by the exchange of virtual charged fermions in a curved space-time.

Specifically, Deser and van Nieuwenhuizen \cite{Deser:1974cz} have attempted to enumerate all possible non-minimal mass dimension 6 actions coupling the Maxwell tensor to the Riemann and Ricci tensors, but excluded parity violating ones because QED is a parity conserving theory. Following that, Berends and Gastmans \cite{Berends:1975ah} computed to one loop order the photon-photon-graviton 3-point correlation function, $\langle T \{ A_\mu A_\nu h_{\alpha\beta} \} \rangle$, for QED in a generic weakly curved spacetime; and later, Drummond and Hathrell \cite{Drummond:1979pp} used their results to do a low energy limit ``matching calculation" to determine exactly the coefficients of the terms obtained by Deser and van Nieuwenhuizen.

With the inclusion of the rest of the electroweak model it is conceivable that the weak interactions would induce, starting at two loops, parity violating mass dimension 6 non-minimal terms; although we are not aware of any explicit calculation to determine their exact coefficients. It cannot be a one loop process because the relevant Feynman diagram would need to contain at least one parity violating current involving the $W$ or $Z$ boson.

Because the action is dimensionless\textsuperscript{\footnotemark[1]}\footnotetext[1]{In this paper, the units $\hbar = c = k_{\text{\tiny B}}=1$ are employed.}, any such additional terms will be suppressed by some inverse power of a mass scale, often times associated with the mass of the virtual particle(s) exchanged.  The parity preserving terms at mass dimension 6 would receive contributions from the standard model beginning at $\mathcal{O}(M_\text{e}^{-2})$ and the parity violating ones possibly starting at $\mathcal{O}(M_\text{W}^{-2})$;  where $M_\text{e}$ and $M_\text{W}$ are the masses of the electron and $W$ boson, respectively.  These dimension 6 terms could also receive contributions from particles that have yet to be observed experimentally, if they are massive enough or if their interactions are sufficiently weak. Therefore, one may hope that constraining the coefficients of such non-minimal terms in the action may in effect probe the existence of new physics. Since these are gravitational interactions, two natural probes are the propagation over cosmological distances of the cosmic microwave background (CMB) photons and solar system tests of General Relativity (GR). Through explicit calculations, however, we will show that one would have to look beyond these avenues to obtain a useful bound.

In this paper, we shall relax the parity preserving assumption of Deser and van Nieuwenhuizen and simply ask what the entire range of possible couplings between the Maxwell tensor and its geometric counterparts is, up to mass dimension 6. As we will see, upon field re-definition, there are only two such terms, so that the most general form of electrodynamics in curved space-time is now given by the action
\begin{align}\label{FinalAction}
&S = -\frac{1}{2} M_\text{pl}^2 \int d^4 x \sqrt{|g|}\( {\cal R} -2\Lambda_\text{cc} \)  - \frac{1}{4}  \int d^4 x \sqrt{|g|}F^{\m\n}F_{\m\n}\notag \\
&+ \int d^4 x \sqrt{|g|}\[ \frac{1}{\L_1^{\phantom{1}2}}F_{\m\n}F_{\a\b}R^{\m\n\a\b} + \frac{1}{2\L_2^{\phantom{2}2}}\widetilde{F}_{\m\n}F_{\a\b}R^{\m\n\a\b}\]
\end{align}
where $M_\text{pl}^2 \equiv (8 \p G)^{-1}$, $F_{\m\n}=\del_{[\m} A_{\n]}=\del_\m A_\n - \del_\n A_\m$ is the Maxwell field tensor, $[\del_\a,\del_\b]V^\m=R^\m_{~~\l\a\b}V^\l$ defines the Riemann tensor, and the dual Maxwell tensor is defined as $\widetilde{F}_{\m\n} \equiv \f{1}{2}\te_{\m\n\a\b}F^{\a\b}$. The $1/\L_1^{\phantom{1}2}$ term has a well-known contribution from QED. While these effects are not new, it is interesting to point out that there is really only one non-trivial term. We can compare this to the work of Drummond and Hathrell \cite{Drummond:1979pp} wherein their coefficient $c/M_{\text{e}}^{~2}= -\a_{\text{\tiny EM}}/(360 \p M_{\text{e}}^{~2})$, where $\a_{\text{\tiny EM}}$ is the fine structure constant, is identified with our $1/\L_1^{\phantom{1}2}$.   As already mentioned, the $1/\L_2^{\phantom{1}2}$ term may be generated by the weak interactions within the standard model.

It is important to note that the mass dimension 6 actions in \eqref{FinalAction} should be viewed as the first terms in an infinite series expansion involving the ratios of the microscopic lengths $1/\L_1$ and $1/\L_2$ to either the wavelength of the photons or that of the characteristic length scale of the gravitation field. That is, the second line of \eqref{FinalAction} is written down with an implicit assumption that the typical energy scale of the photons described by such a theory has to be significantly lower than $\Lambda_1$ and $\Lambda_2$. (One may see this more explicitly by referring to, for instance, the matrix element in equation (2.8) of Drummond and Hathrell \cite{Drummond:1979pp}. The mass dimension 6 contributions to the energy-momentum tensor $\theta^{\mu\nu}$ -- i.e. the terms containing $g_1,g_2$ and $g_3$ -- are $\mathcal{O}(p^2/\Lambda^2)$ or $\mathcal{O}(q^2/\Lambda^2)$ relative to the lowest order mass dimension 4 Maxwell contribution $V_0^{\mu\nu \alpha\beta}$. Here $p$ and $q$ are the momenta of the gravitational field and photon respectively and $\Lambda^2 \sim M_\text{e}^2/\alpha_{\text{\tiny EM}}$.) Just as real electron-positron pairs could be produced if the energy of the photons were of $\mathcal{O}[\text{few MeV}]$, one would have direct access to the (hypothetical) new physics if the energy scale of the photons could reach $\Lambda_1$ or $\Lambda_2$ and the effective theory in \eqref{FinalAction} would then no longer be adequate. At the very least, one would have to include actions up to much higher mass dimensions. As we will see, cosmological and solar system observations are just not sensitive enough to constrain a $\Lambda$ of the same order of magnitude as the photon energies involved ($\sim 3000$K for CMB photons and $\sim 10$ GHz for solar system tests of GR).

In section \ref{sec:II} we list the basic tools needed to construct all the possible non-minimal terms.  Section \ref{sec:III} includes their enumeration from mass dimension two to six, as well as an explanation of why, via a re-definition of the gauge potential $A_\mu$ and the metric tensor $g_{\mu\nu}$, many of these terms are in fact redundant as far as the Maxwell-Einstein system is concerned.  In section \ref{sec:IV} we calculate, for cosmological propagation of CMB light and solar system tests of GR, how accurate observations need to be in order to place useful bounds on $\L_1$ and $\L_2$; they appear impossible to be achieved in the foreseeable future. We conclude and discuss directions for further work in section \ref{sec:V}.

\section{Basic Tools}\label{sec:II}

The most general Lagrangian of gravity and electromagnetism can be written as a sum of the Einstein-Hilbert with a cosmological constant ($\mathcal{L}_{\text{EH},\L_{cc}}$) and the electromagnetic action ($\mathcal{L}_{\text{EM}}$) as defined above, and a perturbative expansion in the mass dimension of the lagrangian density
\begin{equation*}
\int d^4 x \sqrt{|g|}
\(
\mathcal{L}_\text{EH,$\Lambda_{cc}$}
+ \mathcal{L}_\text{EM}
+ \f{1}{M_\star}{\cal L}_5 + \f{1}{M_\star^2}{\cal L}_6
+ \dots
\)
\end{equation*}
with $M_\star$ representing the lowest of the (possibly many) mass scales that are physically relevant.

In order to systematically enumerate all the possible non-minimal terms coupling gravity to electromagnetism, we start by listing the most rudimentary tensors available, before forming the general set of scalars out of them. The requirements of U(1) gauge invariance and general coordinate covariance lead us to the {\it dimensionful} tensors, the covariant derivative and field strengths
\begin{equation}\label{dimful tensors}
\del_\m, ~F_{\m\n},  ~R_{\m\n\a\b}
\end{equation}
with mass dimension 1, 2, and 2 respectively.  The primitive {\it dimensionless}\textsuperscript{\footnotemark[2]}\footnotetext[2]{Strictly speaking, the physical dimensions of different components of a given tensor are the same when computed in an orthonormal frame, where the metric, in particular, is then $g_{\mu\nu} \to \eta_{\mu\nu} \equiv \text{diag}[1,-1,-1,-1]$. However, since we are forming scalars out of these tensors -- the result is independent of whether the coordinate or orthonormal frame was chosen -- we are delineating this construction within a coordinate frame, where calculations are easier.} geometric objects are
\begin{equation}\label{dimless tensors}
g^{\m\n}, ~\widetilde{\ep}^{\m\n\a\b} \equiv \f{1}{\sqrt{|g|}}\ep^{\m\n\a\b}
\end{equation}
where $\ep^{\m\n\a\b}$ is the fully anti-symmetric Levi-Civita symbol, and we define $\ep^{0123} \equiv -1$.  Scalars built from the covariant Levi-Civita tensor will violate parity since it transforms as a pseudo-tensor (even under parity). As far as the complete enumeration of the primitive tensors is concerned, the placement of their indices (upper or lower) is immaterial because we will be forming scalars out of them anyway.  The basic strategy for constructing the most general set of scalars is, then, to consider all possible contractions between appropriate products of \eqref{dimful tensors} and \eqref{dimless tensors}. We need not consider derivatives on the $\widetilde{\epsilon}$-tensor as $\nabla_\tau \widetilde{\epsilon}^{\mu\nu\alpha\beta} = 0$. Furthermore, because
\begin{equation}
\te^{\m\n\a\b}\te_{\r\t\s\l}= -\de^{\m}_{[\r} \de^{\n}_\t \de^{\a}_\s \de^{\b}_{\l]}
\end{equation}
we see that it suffices to consider terms that contain zero or one $\widetilde{\epsilon}$-tensor only.

\section{Enumeration of terms}\label{sec:III}
Schematically, we seek to form combinations of the type
\begin{align}\label{Schematic}
\del^a F^b R^c
\end{align}

As $F_{\m\n}$ and $R_{\m\n\a\b}$ are both of dimension $\[M\]^2$, we look for terms satisfying $a+2b+2c \leq 6$.  As all tensors except $\del_\m$ have an even number of indices, in order to form a scalar we see that  $a$ must be an even number,  therefore no odd mass dimension terms exist. Since we are considering couplings between both the electromagnetic and gravitational field, we require $b > 0$. We have summarized these possibilities in Table \ref{tab:table1}.

\begin{table}
\begin{ruledtabular}
\begin{tabular}{m{2cm}m{7cm}}
Mass Dimension & Possible Combinations of (a,b,c) in \eqref{Schematic}\\
\hline
2 & (0,1,0)\\
4 & (0,1,1), (0,2,0), (2,1,0)\\
6 & (0,1,2), (0,2,1), (0,3,0), (2,1,1), (2,2,0),  (4,1,0)
\end{tabular}
\end{ruledtabular}
\caption{\label{tab:table1} The possible combination of tensors at various dimensions.  }
\end{table}
As we enumerate the possible terms, we  will
 not consider trivial numerical factors as these would
be absorbed into coefficients of the Lagrangian anyway.  The following Bianchi identities are quite useful
\begin{eqnarray}
\del_{[\a}F_{\m\n]}=0\label{DF} \\
R_{\m[\n\a\b]}=0\label{R_Bianchi}\\
\del_{[\n} R_{\l\s]\r\t}=0 \label{DR_Bianchi}
\end{eqnarray}
Using \eqref{R_Bianchi} it is not hard to show that
\begin{equation}
\te^{\m\n\r\t}R_{\m\n\l\s}= 2 \te^{\m\n\r\t}R_{\m\l\n\s} \label{Ident_ep_R=2ep_R}
\end{equation}
The anti-symmetric nature of $F^{\mu\nu}$ implies
\begin{align}\label{Identity3}
\del_\m \del_\n F^{\m\n} = R_{\l\n}F^{\l\n} =g^{\mu\nu} F_{\mu\nu} = 0
\end{align}
In 4 dimensional spacetimes, the antisymmetrization of more than 4 indices always yields zero. One useful corollary \cite{Giannotti:2008cv} is
\begin{equation}\label{corollary}
g^{\a[\b}\widetilde{\epsilon}^{\m\n\r\t]}=0
\end{equation}

\ssec{Dimension: 2 and 4}
At dimension 2, the only scalar that can be formed here is $F_{\m\n}g^{\m\n}$, which is zero.  Next we consider the dimension 4 terms $F_{\m\n}R_{\a\b\g\de}$, $F_{\a\b}F_{\m\n}$, and $\del_\m\del_\n F_{\a\b}$, which we call type I, II and III, respectively.  In addition to contractions with the metric, we must separately consider both contractions without and with the Levi-Civita tensor.
\sssec{Without Levi-Civita Contractions}
{\bf Type I}: The only scalar is $F^{\mu\nu} R_{\mu\nu} = 0$.

{\bf Type II}:  The scalar this forms is the one found in the canonical Maxwell Lagrangian
\begin{align*}
F_{\m\n}F^{\m\n}
\end{align*}

{\bf  Type III}: This is not only a total derivative, which does not contribute to the dynamics, it is also identically zero from \eqref{Identity3}.

\sssec{With Levi-Civita Contractions}
{\bf  Type I}: Because of \eqref{R_Bianchi}, at most 2 indices on the Riemann tensor may contract with the Levi-Civita, leaving only the possibility of $F_{\m\n}R_{\a\b}\te^{\m\n\a\b}$, which is zero due to the symmetry of the Ricci tensor.

{\bf Type II}: The only scalar that can be formed here is $F_{\m\n}F_{\a\b}\te^{\m\n\a\b}$. Using identity \eqref{DF}, we see this is a total derivative:
\begin{eqnarray}
\del_\m\(\te^{\m\n\a\b}A_\n F_{a\b}  \)&=& \widetilde{F}^{\alpha\beta} F_{\alpha\beta}
\end{eqnarray}

{\bf Type III}: Zero, by the Bianchi identity \eqref{DF}.

\ssec{Dimension: 6}
Here we consider the terms (A) $F_{\m\n}R_{\a\b\g\de}R_{\r\t\s\l}$, (B) $F_{\m\n}F_{\r\t}R_{\a\b\g\de}$, (C) $F_{\m\n}F_{\a\b}F_{\r\t}$, individual terms from (D) $\del_\r\del_\t F_{\m\n}R_{\a\b\g\de}$ and (E) $\del_\m\del_\n F_{\a\b}F_{\r\t}$, and (F) $\del_\m\del_\n \del_\r \del_\t F_{\a\b}$. We again separately consider contractions without and with the Levi-Civita tensor.  As illustrated in the previous section, there are many tricks that simplify such constructions greatly.  While we will not discuss all of them, we have tried to give the most salient  examples.
 We have therefore only listed the non-zero, non-redundant terms below.
\R
\sssec{Without Levi-Civita Contractions}

{\bf Type A}: All of these are zero because while $F_{\mu\nu}$ is antisymmetric, all two-index tensors built from contracting two Riemann tensors are symmetric in the indices. As an example, using \eqref{R_Bianchi} it is possible to show that
\begin{align}
R_\m^{~\r\b\t}R_{\n\b\r\t} = \f{1}{2}R_\m^{~\r\b\t}R_{\n\r\b\t}
\end{align}
where $R_\m^{~\r\b\t}R_{\n\r\b\t}$ is $(\m\leftrightarrow\n)$ symmetric.

{\bf Type B}:
The non-redundant terms are
\begin{align}\label{non minimal 1}
&F_{\m\n}F^{\m\n}{\cal R}\notag\\
&F^{\m}_{~\a}F_{\m\b}R^{\a\b}\notag\\
&F_{\a\b}F_{\m\n}R^{\a\b\m\n}
\end{align}

{\bf Type C}: All of these are zero due to symmetry considerations.

{\bf Type D}:  These terms are either equivalent to type A (which are zero) or are of the type ~$\del^\m F_{\m\n}V^\n$  (for some appropriate $V^\n$), which can all be made to vanish at this order in mass dimensions by a suitable field re-definition, $A^\n \to A^\n + \de A^\n$, where $\de A^\n \propto  V^\n$.  While it is true that this field re-definition will induce a variation of all the mass dimension 6 terms, the new terms appearing are of even higher order, and so are irrelevant for this analysis.

{\bf Type E}:  As in \cite{Deser:1974cz}, one can show using \eqref{DF} that, up to a total derivative, there is an equivalence between the term
\begin{align*}
\(\del_\a F_{\b\g}\)\(\del^\a F^{\b\g} \)
\end{align*}
and
\begin{equation*}
2\(\del_\a F^{\a}_{~\g}\)\(\del_\b F^{\b\g}\) + 2F^{\a}_{~\g}F^{\g\l}R_{\l\a} + F^{\a\b}F^{\l\g}R_{\a\b\g\l}
\end{equation*}
Therefore, the non-zero terms here are equivalent to type B and D.

{\bf Type F}: This is a total derivative, and does not contribute to the dynamics.

 \sssec{With Levi-Civita Contractions}
{\bf  Type A}:  Zero, for the same reasons as above.  To show this rigorously, \eqref{corollary} may be useful.

{\bf  Type B}: As an example, using \eqref{corollary} it is possible to show
\begin{align}
&F_{\a\l}F_{\b\t}R_{\m\n}^{~~\a\b}\te^{\m\n\l\t}=R_{\m\n}^{~~\a\l}F_{\a\l}\widetilde{F}^{\m\n} - 2R_{\m\a}\widetilde{F}^{\m\l}F^\a_{~\l}
\end{align}
In all, the non-redundant terms here are
\begin{align}\label{non minimal 2}
&F_{\m\n} \widetilde{F}^{\m\n}{\cal R}\notag \\
&F_{\m}^{~\s} \widetilde{F}^{\m\n} R_{\n\s}\notag\\
&\widetilde{F}^{\a\b}F^{\r\s}R_{\a\b\r\s}
\end{align}

{\bf Type C}: Zero or redundant, as above.

{\bf Type D}: Zero, as above.

{\bf Type E}: These are either redundant or zero, as above.  Consider the term  $\te^{\m\n\a\b} \del_\m F_{\n\s}\del_\a F^\s_{~\b}$.  After integration by parts, we find it proportional to
\begin{align}
&\del_\a \( \te^{\m\n\a\b} \del_\m F_{\n\s}\) F^\s_{~\b} \notag \\
&\qquad = \frac{1}{2} \te^{\m\n\a\b} F^\s_{~\b} (R^{~\l}_{\n~\a\m}   F_{\l\s} + R^{~\l}_{\s~\a\m}F_{\n\l})
\end{align}

{\bf Type F}: This is a total derivative, and does not contribute to the dynamics.

At this stage, there are six possible non-minimal coupling terms; three \eqref{non minimal 1} of which are consistent with those listed by Drummond and Hathrell \cite{Drummond:1979pp}, while the other three \eqref{non minimal 2} are new, and are their parity-violating counterparts.

\ssec{Absorption of Terms via Metric Re-definition}

Up to this point, the general action is of the form
\begin{align}\label{GeneralAction}
S = S_{\text{EH},\L_\text{cc}} + S_{\text{EM}} + \int d^4 x \sqrt{|g|}\[ A^{\m\n}R_{\m\n} + B {\cal R} +\ \dots\]
\end{align}
where $(\dots)$ indicates terms that contain neither the Ricci tensor nor scalar (but may contain the Riemann tensor).  We will now demonstrate that $A^{\mu\nu}$ and $B$ can be eliminated through a suitable change of variables, i.e. a re-definition of the metric.  We make the transformation, $g^{\a\b} \to g^{\a\b} + \de g^{\a\b}$ and choose
\begin{equation}
\de g^{\a\b} = - \f{2}{M_\text{pl}^2}\(\l g^{\a\b} + f^{\a\b}   \)
\end{equation}
where $\l$ and $f^{\a\b}$ obey the relations
\begin{align}
f_{\a\b} = -A_{\a\b}, \qquad
\l + \frac{f}{2} = B
\end{align}
where $f$ is the trace of $f_{\a\b}$. The overall modification to the action \eqref{GeneralAction} is then
\begin{align}\label{NewAction}
S&= S_{\text{EH},\L_\text{cc}} + S_{\text{EM}} + \de S + \dots\\
\de  S &\equiv \int d^4 x \sqrt{|g|}\L_\text{cc} g^{\m\n}\( A_{\m\n} + B g_{\m\n}\)
\end{align}
where the ($\dots$) represent the same omitted terms.  While it is true that the change of variables will affect all terms, it may be checked that the variation to the non-minimal terms occurs at least at mass dimension 8, and so will not have an impact on this analysis.

Discarding the total derivative $\widetilde{F}_{\m\n}F^{\m\n} $, we see that
\begin{align}\label{NewPiece}
\de S \to -\f{1}{4} \int d^4 x \sqrt{|g|} \f{\b \L_\text{cc} }{M_\star^2}  F_{\m\n}F^{\m\n}
\end{align}
for an appropriate mass scale, $M_\star$ and constant, $\b$.  This re-definition could pose an issue if it were to flip the overall sign of the Maxwell action, but will not be a problem as long as $M_\star^2 \gg \L_\text{cc}$.  This modification to the canonical Maxwell Lagrangian can be absorbed by simply rescaling the electromagnetic potential as
\begin{equation*}
 A_\m \to A_\m \(1+ \f{ \b \L_\text{cc}}{M_\star^2} \)^{-\f{1}{2}}
\end{equation*}
Upon these field re-definitions, we obtain the advertised result \eqref{FinalAction}.

\section{Observables and Constraints}\label{sec:IV}
Varying the action \eqref{FinalAction} with respect to $A_\m$, the modified Maxwell equations are
\begin{align}
\nabla^\mu F_{\mu\nu} &+ \frac{1}{\Lambda_1^2}\( \nabla_{[\alpha} R_{\beta]\nu} + R_{\mu\nu \alpha\beta} \nabla^{\mu} \) F^{\alpha\beta}\notag \\
\qquad& + \frac{1}{4 \Lambda_2^2}
\( R_{\rho\sigma \alpha\beta} \widetilde{\epsilon}^{\rho\sigma \mu}_{\phantom{\rho\sigma \mu} \nu} \nabla_{\mu} F^{\alpha\beta}
+ 2 \widetilde{F}^{\rho\sigma} \nabla_{[\rho} R_{\sigma]\nu}\right.\notag\\
&\left.+ 2 R^\mu_{\phantom{\mu}\nu\rho\sigma} \nabla_\mu \widetilde{F}^{\rho\sigma} \) = 0\label{Maxwell}
\end{align}

Within the cosmological context, we will derive the general solutions to \eqref{Maxwell} in a spatially flat Friedmann-Robertson-Walker (FRW) metric and examine their implications for the propagation of the cosmic microwave background. We will then see, as also discussed by \cite{Prasanna:2003ix}, although one might have hoped that the large distances involved would help accumulate effects from these non-minimal terms and render them discernible, the Hubble parameter of our universe is simply too small for cosmology to be a sensitive probe.

We shall also show below that these non-minimal terms do modify the geodesics followed by photons in a curved background, defining an effective metric, so that the travel time and deflection angles of light about massive objects will be altered from their standard values. Observations of the Shapiro delay of radio signals from the satellite Cassini currently provide the tightest bound on the PPN parameter $\gamma_\text{\tiny PPN}$. Even though these Cassini observations yield tighter restrictions on $\L_{1,2}$ than the cosmological ones, they still lie significantly below the threshold necessary to provide physically useful constraints.

\ssec{Cosmological Constraints}

We start first with cosmological probes and work with a spatially flat FRW universe, where $g_{\mu\nu} = a(\eta)^2 \eta_{\mu\nu}$.  We then proceed to solve, in the Coulomb gauge ($A_0 = 0$), the general solutions of the vector potential $A_\mu$ to the vacuum modified wave equations, using the JWKB approximation. To this end, if ${\bf k}$ is the spatial momentum vector of the photon, it helps to expand the spatial portion of $A_\mu$, ${\bf A}$, in terms of basis vectors where one of them is parallel to ${\bf k}$ and the other two correspond to left- and right-circular polarizations.  That is, if we first consider an orthonormal basis defined by unit vectors, $\{\hat{\bf e}_\text{I}, \hat{\bf e}_\text{II}, \f{{\bf k}}{|{\bf k}|}\}$, then define
\begin{eqnarray}
{\bf A\(\e,{\bf x}\)} \equiv \(A_+\(\e\) \hat{\bf e}_+ + A_-\(\e\) \hat{\bf e}_-\)e^{- i {\bf k}\cdot {\bf x}}
\end{eqnarray}
where
\begin{align}
\hat{\bf e}_{\pm} \equiv \f{1}{\sqrt{2}}\(\hat{\bf e}_\text{I} \pm i\hat{\bf e}_\text{II}\)
\end{align}
the resultant equation of motion is (from \eqref{Maxwell})
\begin{align}\label{Apm}
A_\pm'' \(1+ \psi \) + A_\pm' \psi' + {k}^2 A_\pm \(1 + \chi \) \pm 2 \phi' k A_\pm = 0\notag \\
\end{align}
Here $k \equiv |{\bf k}|$, the prime denotes derivatives with respect to conformal time, $\e$, and we have used the following definitions
\begin{align}
\psi(\e) &\equiv -\frac{2}{\L_1^2}\[ \f{a''}{a^3}   -\(\frac{a'}{a^2}\)^2  \] \\
\chi (\e) &\equiv -\frac{2}{\L_1^2}\(\frac{a'}{a^2}\)^2   \\
\phi(\e) &\equiv \f{1}{2}\frac{1}{\L_2^2} \frac{a''}{a^3}
\end{align}

Following \cite{Carroll:1991zs}, we now attempt a JWKB solution by first requiring the solutions take the form
\begin{align}
A_\pm (\e) &= ~{\cal{A}}_\pm \exp{ \[ i \int^\e_{\e_0} d\e' f_\pm(\e')\]}\label{WKB}
\end{align}
We next assume that time derivatives of ${\cal{A}}$ are negligible, $f' \ll f^2$, and proceed to insert \eqref{WKB} into \eqref{Apm}.   As they are small for the cosmological eras of interest, we expand $f$ to linear order in $\phi, \chi \text{ and }\psi$ to find (choosing a positive root)
\begin{equation}
f \approx k +i\frac{\psi'}{2} + \frac{1}{2}k\(\chi - \psi \) \pm \phi'
\end{equation}
so that
\begin{align}
A_\pm(\e) &\approx {\cal{A}} ~\exp \bigg[ ik\(\Delta\eta + \frac{1}{2} \int^\e_{\e_0} d\e' \(\chi - \psi \)\) \\
&\qquad \qquad \qquad -\frac{\Delta\psi}{2}  \pm i  \Delta\phi\bigg]\notag
\end{align}

The $i k \Delta\eta$ is just the usual plane wave term. The parity-conserving $1/\L_1^{\phantom{1}2}$ term contributes to the phase a real part, the integral of $\chi-\psi$, and a dissipative imaginary part, $i\Delta\psi/2$. The birefringent $\pm \Delta\phi$ arises from the parity-violating $1/\L_2^{\phantom{1}2}$ term.

We may now write
\begin{equation}
A_\pm \propto \exp\left[i\th_\pm - \f{1}{2}\Delta \psi\right],
\end{equation}
with the same proportionality holding for the two circular polarizations of the electric field, $E_\pm$. Hence, the energy density of electromagnetic waves propagating through the universe will be suppressed by a factor $e^{-\Delta \psi}$. The QED contribution can be obtained by borrowing Drummond and Hathrell's result \cite{Drummond:1979pp}, that tells us that $1/\L_1^{\phantom{1}2} \approx -10^{-3}\a_{\text{\tiny EM}}/M_\text{e}^{~2}$, which leads us to find an extremely small damping of roughly  $ \exp{[-10^{-73}]}$, if QED is the most dominant contribution.

For light coming from a linearly polarized source, over the course of its propagation the plane of polarization rotates by an angle, $|\Delta \a| = \f{1}{2}\(\th_+ - \th_-  \)= \abs{\Delta \phi}$.  Since (in observer time) we have $\phi(z) \sim (H(z)/\L_2)^2$ and during the matter-dominated era, $H^2 \propto  (z+1)^3$, bounding the observed rotation angle restricts the mass scale as
\begin{equation}
\L_2 \gtrsim H_0 \( \f{(z+1)^{3}}{\abs{\Delta \a}} \)^{\f{1}{2}}, \label{biref}
\end{equation}
where $H_0 \approx 2 \times 10^{-33}$eV.

While there are astrophysical ($0 < z < 4$) sources of polarized radiation, such as radio galaxies (see \cite{Alighieri:2010eu} for example) which provide ${\cal O}(1^\circ)$ limits  on polarization rotation, the CMB turns out to give the best constraint because of its large redshift ($z \approx 1100$). By rotating the plane of linear polarization, birefringence mixes the E and B polarization modes of CMB. Specifically, this induces a non-zero cross correlation between the temperature anisotropy and B modes, given by $C^{TB}_\ell = C^{TE}_\ell \sin(2 \Delta \a)$ \cite{Cabella:2007br}.  WMAP \cite{Komatsu:2010fb} has put ${\cal O}(1^\circ)$ limits on the (isotropic) rotation angle of linear polarization of the CMB. By applying \eqref{biref}, we obtain a naive constraint of $\L_2 \gtrsim 10^{-33} \text{MeV}$. However, one must remember that the energy of the CMB photons themselves are of $\mathcal{O}(3-3000)$K (or, $\mathcal{O}(3 \times 10^{-10} - 3 \times 10^{-7})$ MeV), at least 23 orders of magnitude greater than this lower bound. To obtain a physically meaningful bound, one ought to ask instead, how accurate does $\Delta \alpha$ need to be determined for a $\L_2$ of at least the same energy scale as that of the photons? Setting $\L_2 \gtrsim 3000$K, and inverting the inequality \eqref{biref}, the answer is $\Delta \alpha \lesssim \mathcal{O}(10^{-55})$. It should be safe to assume this is out of observational reach.

Although this effect is miniscule, it is interesting to find a possible standard model source of cosmological birefringence that does not invoke any new degrees of freedom or extra dynamics (e.g. \cite{Ni:1977, Carroll:1989vb}).

\ssec{Solar System Constraints}

Next, we would like to examine how our mass dimension 6 terms modify standard GR predictions, so that we may use observations to constrain their coefficients. In the following, we will employ the JWKB approximation to work out the photon's modified dispersion relations, due to the addition of the non-minimal terms to Maxwell's equations. We then extract the effective metric experienced by these photons, and compute the induced corrections to both deflection angle and the Shapiro delay of light propagating past a massive body.

Let us first consider the modified Maxwell's equations with an example background geometry given by the Schwarzschild metric:
\begin{equation}
d\tau^{2}=B(r)dt^{2}-A(r)dr^2-r^{2}d\theta^{2}-r^{2}\sin\theta^{2}d\phi^{2}\label{Smetric}
\end{equation}
where $B(r)=U(r)$, $A(r)=U^{-1}(r)$, with $U(r)=1-\frac{r_{s}}{r}$, $r_{s}=2G M$, and $M$ is the mass of the object. Note that we have neglected the cosmological constant, $\L_{cc}$.

Since the wavelength of the light considered here is much smaller than the background metric's radius of curvature, we will use the JWKB ansatz, $A_{\mu} = \text{Re}(a_{\mu}e^{i\psi})$, in which the amplitude $a_{\mu}$ is slowly varying while the phase $e^{i\psi}$ varies rapidly.  Under these consideration, the modified Maxwell equations (\ref{Maxwell}) become
\begin{align}
0 &= \bigg( k_{\mu}k^{\mu}\delta^{\nu}_{\beta}-\frac{2}{\Lambda_{1}^{2}}R^{\nu}_{\phantom{\nu}\mu\alpha\beta}k^{\mu}k^{\alpha} \nonumber\\
&\qquad + \frac{1}{2\Lambda_{2}^{2}}\left(
R_{\rho\sigma\alpha\beta} \widetilde{\epsilon}^{\rho\sigma\mu\nu} k_{\mu}k^{\alpha}
- R^{\nu}_{\phantom{\nu}\mu\rho\sigma} \widetilde{\epsilon}^{\rho\sigma\alpha}_{\phantom{\rho\sigma\alpha}\beta} k^{\mu}k_{\alpha} \right) \bigg) a^{\beta} \nonumber\\
&\equiv N^{\nu}_{\phantom{\nu}\beta}a^{\beta}
\end{align}
For the system of equations to have non-trivial solutions, we require $\det N^{\nu}_{\phantom{\nu}\beta} = 0$. The eigenvalues of $N$ give us the photon's dispersion relations, and the corresponding null eigenvectors are the polarization vectors.

To simplify the algebra, it is helpful to rewrite our equations in an orthonormal basis using the vierbeins $e^{\h{b}}_{\phantom{b}\beta}$, defined as
\begin{equation}
g_{\m\n}=e^{\h{a}}_{~\m}e^{\h{b}}_{~\n}\e_{\h{a}\h{b}}
\end{equation}
so that
\begin{eqnarray}
k_{\h{b}}&=&k_{\nu}e^{\phantom{b}\nu}_{\h{b}}\\
N^{\h{c}}_{\phantom{c}\h{b}}&=&e^{\h{c}}_{\phantom{c}\nu}N^{\nu}_{\phantom{\nu}\beta}e^{\phantom{b}\b}_{\h{b}}
\end{eqnarray}

Our conventions here reserve the greek indices for the coordinate frame ($t, r, \theta, \phi$) and latin for the orthonormal frame ($\h{t}, \h{r}, \h{\th}, \h{\ph}$). In matrix form, the vierbeins are
\begin{equation}
e^{\h{b}}_{~\mu} = \(\begin{array}{cccc}
\sqrt{U} & 0 & 0 & 0 \\
0 & 1/\sqrt{U} & 0 & 0\\
0 & 0 & r & 0\\
0 & 0 & 0 & r\sin \theta
\end{array} \)
\end{equation}
and $e^{\phantom{b}\m}_{\h{b}}$ is just the inverse of $e^{\h{b}}_{~\mu}$, i.e. $e^{\phantom{b}\m}_{\h{b}}e^{\h{b}}_{~\nu}=\delta^{\mu}_{~\nu}$, and $e^{\h{a}}_{~\mu}e^{\phantom{b}\m}_{\h{b}}=\delta^{\h{a}}_{~\h{b}}$. Due to the spherical symmetry of the Schwarzschild metric, we can consider, without loss of generality, the light propagation to lie in the $\theta=\frac{\pi}{2}$ plane, i.e. $k_{\h{\th}} = 0$. From here on we also choose, for simplicity, to assume that the two hypothetical energy scales are the same, namely, $\L_1=\L_2\equiv\L$.
Under these considerations, we find 
\begin{equation}
N^{\h{b}}_{~\h{c}} = \( \begin{array}{cccc}
\Sigma+\Delta_{11}& -\Delta_{12} & -\Delta_{13}&  -\Delta_{14} \\
\Delta_{12} & \Sigma-\Delta_{22} & \Delta_{23} & \Delta_{24}\\
\Delta_{13} & \Delta_{23} & \Sigma+\Delta_{33}& 0 \\
\Delta_{14}&  \Delta_{24} & 0 & \Sigma+\Delta_{44}
\end{array} \)
\end{equation}
where
\begin{align}
&\Delta_{11}=\frac{(2k_{\h{r}}^2-k_{\h{\ph}}^{2})r_{s}}{r^{3}\Lambda^{2}} &  &\Delta_{22}=\frac{(2k_{\h{t}}^{2}+k_{\h{\ph}}^{2})r_{s}}{r^{3}\Lambda^{2}}\notag\\
&\Delta_{33}=\frac{(k_{\h{t}}^{2}-k_{\h{r}}^{2}+2k_{\h{\ph}}^{2})r_{s}}{r^{3}\Lambda^{2}}&    &\Delta_{44}=\frac{(k_{\h{t}}^{2}-k_{\h{r}}^{2})r_{s}}{r^{3}\Lambda^{2}}\notag\\
&\Delta_{12}=-\frac{2k_{\h{t}}k_{\h{r}}r_{s}}{r^{3}\Lambda^{2}}&  &\Delta_{13}=\frac{3k_{\h{r}}k_{\h{\ph}}r_{s}}{r^{3}\Lambda^{2}}\notag\\
&\Delta_{14}=\frac{k_{\h{t}}k_{\h{\ph}}r_{s}}{r^{3}\Lambda^{2}}&   &\Delta_{23}=\frac{3k_{\h{t}}k_{\h{\ph}}r_{s}}{r^{3}\Lambda^{2}}\notag\\
&\Delta_{24}=\frac{k_{\h{r}}k_{\h{\ph}}r_{s}}{r^{3}\Lambda^{2}}
\end{align}
and $\Sigma=k_{\h{t}}^{2}-k_{\h{r}}^{2}-k_{\h{\ph}}^{2}$.

Two of the eigenvalues of $N^{\h{b}}_{~\h{c}}$ are $k_{\h{t}}^{2}-k_{\h{r}}^{2}-k_{\h{\ph}}^{2}=0$; this is the canonical light-cone dispersion relation. However, their corresponding polarization vectors are pure-gauge modes, and thus non-physical. The other two, to $\mathcal{O}(\L^{-2})$, are
\begin{eqnarray}\label{modified dispersion}
k_{\h{t}}^{2}-k_{\h{r}}^{2}-k_{\h{\ph}}^{2}\(1\pm \frac{3\sqrt{2}r_{s}}{\L^2 r^3} \)=0
\end{eqnarray}

At this order, if $\L_1$ and $\L_2$ had been kept distinct, the non-minimal modifications to the dispersion relations add in quadrature and are thus symmetric under the interchange of $\L_1$ and $\L_2$.  Therefore, any physical effects derived from these relations will not distinguish between the two; however, this symmetry does not hold for their corresponding polarization vectors.

\sssec{Effective metric solution}

We wish to analyze the implications of this modified dispersion relation in terms of a modified metric, following Myers and Lefrance  \cite{Lafrance:1994in}. The two dispersion relations \eqref{modified dispersion} could be viewed as $k_{\h{a}} k_{\h{b}} \widetilde{g}^{\h{a}\h{b}}=0$ or, in a coordinate frame, $k_\mu k_\nu \widetilde{g}^{\mu\nu}=0$.  This defines the effective metric, $\widetilde{g}_{\m\n}$, for each of the two dispersion relations, however, only up to an overall conformal factor.  We choose to write it as
\begin{equation}\label{metric}
d\tau^{2}={\cal B}(r)dt^{2}-{\cal A}(r)dr^{2}-r^{2}d\phi^{2}\\
\end{equation}
where
\begin{align}\label{metric defs}
{\cal B}(r)&\equiv(1\pm\delta)\left(1-\frac{r_{s}}{r}\right)\\
{\cal A}(r)&\equiv(1\pm\delta)\left(1+\frac{r_{s}}{r}\right)\\
\delta&\equiv\frac{3\sqrt{2}r_{s}}{\Lambda^{2}r^{3}}
\end{align}

To $\O{r_s}$, the modification to the dispersion relations amount to a re-scaling of $B(r)$ and $A(r)$ found in \eqref{Smetric} by a factor of $1\pm \delta$, as predicted by the standard result in GR.  Notice that we continue to work in the $\th=\f{\p}{2}$ plane without loss of generality. Since there is still no time dependence, the metric also remains static.

In this modified geometry the contravariant wave vector, $k^{\m}=\widetilde{g}^{\m\n}k_{\n}$, is the tangent vector to the path normal to the surfaces of constant phase $\psi$, i.e. $k^\mu = dx^\mu/ds$, where $s$ is an appropriate affine parameter.  In order to be convinced that is the right interpretation, all that is needed is to show that the modified dispersion relation implies that $x^\m(s)$ is a geodesic of this modified spacetime.

We show this, following  \cite{MTW:1973}, by first taking a covariant derivative of the dispersion relation, now with respect to the effective metric
\begin{equation}
\widetilde{\del}_\a \(k_\m k_\n \widetilde{g}^{\m\n} \)= 2 k^\m \widetilde{\del}_\a k_\m = 0
\end{equation}
Since $k_\m=\widetilde{\del}_\m \psi=\d_\m \psi$, it is straightforward to show that
\begin{equation}
\widetilde{\del}_\a k_\m = \widetilde{\del}_\m k_\a
\end{equation}
Thus
\begin{equation}
k^\m \widetilde{\del}_\m k_\a =  0 = k^\m \widetilde{\del}_\m k^\a
\end{equation}
which is none other than the geodesic equation.

Though this effective metric is defined only up to an overall conformal factor, if one were to multiply it by any function of $r$, the trajectory of the null geodesics remains unaltered. Therefore, the predictions we will quote below for the deflection angle and modified Shapiro delay which are calculated based on (\ref{metric}) are unambiguous.

\sssec{Deflection Angle}

For a general metric of the form \eqref{metric}, the total deflection angle of light passing by a massive object is (see, e.g. \cite{weinberg})
\begin{eqnarray}
\Delta\Phi&=&2\int_{r_0}^{\infty}{\frac{dr}{r}\sqrt{\frac{A(r)}{\frac{B(r_0)r^{2}}{B(r)r_0^{2}}-1}}}-\pi\label{deltaphi}
\end{eqnarray}
where $r_0$ is taken to be the point of closest approach of the light from the object. Expanding the integrand in powers of $r_{s}/r$, the integral above gives us
\begin{equation}
\Delta\Phi=2\(\frac{r_{s}}{r_0}\pm\frac{2\sqrt{2}r_{s}}{\Lambda^{2}r_0^{3}}\)\label{dphi}
\end{equation}

The first factor is just the standard contribution from the Schwarzschild metric, and the second term is the new contribution of the non-minimal terms.   An unpolarized light ray traversing close to the massive body would incur a splitting due to the opposite signs arising in the form of \eqref{dphi}.  While there are observational limits on the deflection angle of light passing close to our sun, these limits are not as stringent as those derived in the following section, and we will therefore not purse a constraint on our non-minimal terms here. The gravitational deflection of light computation here suggests that our non-minimal terms would also modify the weak lensing signals currently sought by large scale structure observations.

\sssec{Modified Shapiro Time Delay}

We now move on to consider the time-of-flight of a null light ray propagating between two points in space such that it passes close to a massive object. Such a light ray is known to experience a delay in its time of flight, relative to the same flight in Euclidean space. This is commonly referred to as Shapiro delay \cite{Shapiro:1964uw}.

For ease of comparison to and discussion in reference to the literature, we will now switch to the
isotropic gauge in calculating the modification to this delay.  In this coordinate system, the Schwarzschild metric \eqref{Smetric} is written as
\begin{align}\label{newmetric}
d\t^2=B(r) dt^2  -A(r)&\(dr^2 + r^2d\th^2 +r^2\sin^2{\th}d\ph^2\)
\end{align}
where
\begin{align}
 B(r)&=\(\f{1-\f{r_s}{4r}}{1+\f{r_s}{4r}}\)^2\\
 A(r)&=\(1+\f{r_s}{4r}\)^4
\end{align}

To first order in $r_s$, both the matrix $N^{\h{b}}_{~\h{c}}$ and the form of the modified dispersion relations remains unchanged. The expressions that do change are the associated vierbeins and the effective metric. In particular, the latter becomes
\begin{align}
d\t^2=\(1\pm\de\)B(r) dt^2  -\(1\pm\de\)A(r)dr^2 \nonumber\\
-A(r)\( r^2d\th^2 +r^2\sin^2{\th}d\ph^2\) \label{Shapiro Delay}
\end{align}
where $\de$ was defined in \eqref{metric defs}. By approximating the null path to be a straight line in space, the delay in the round-trip time-of-flight between the two points ($P_{1,2}$) is
\begin{equation}\label{Modified Delay}
\Delta t=   2r_s \log{\[\f{4X_2 X_1}{b^2}\]} \pm \f{4\sqrt{2}r_s}{b^2 \L^2}
\end{equation}
If $Q$ is the point on the straight line joining $P_1$ and $P_2$ closest to the massive object, then $b$ is the distance between $Q$ and the object and $X_{1,2}$ are the distances from $Q$ to $P_{1,2}$, respectively.  We note that $b$ is not equal to the actual distance of closest approach, $r_0$, used in \eqref{deltaphi}. A discussion of this and the different ways of calculating $\Delta t$ that appear in the literature may be found in appendix \eqref{app}.

Within the Parametrized Post-Newtonian (PPN) formulation, the Schwarzschild metric is altered to quantify deviations from GR \cite{Will:2005va}. To date, the most precise measurement of the parameter $\gamma_\text{\tiny PPN}$, which is equal to $1$ in GR, comes from the observation of the Shapiro delay from the Cassini spacecraft \cite{Bertotti:2003rm}. Under this parametrization, and in the idealized limit where the Earth and Cassini are stationary, the Shapiro delay is
\begin{equation}\label{PPN delay}
\Delta t=  \(1+\gamma_\text{\tiny PPN}\)r_s \log{\[\f{4X_2 X_1}{b^2}\]}
\end{equation}
The Cassini experiment measured the fractional Doppler-frequency shift of the radio carrier waves, which in turn is the time derivative of the Shapiro time delay $y(t) = d\Delta t/dt$. Since the most rapidly changing length scale in \eqref{PPN delay} is the straight-line closest approach distance, $y(t) \propto db/dt$.

In order to put a constraint on the energy scale $\L$, we set $y(t)$ as determined by \eqref{Modified Delay} equal to $y(t)$ as determined by \eqref{PPN delay}, thus determining the lower bound on $\L$. The actual interpretation and calculation involved for the timing measurement is quite involved and the reader is referred to \cite{moyer} for further information on the details on the actual treatment.  Using the experimental parameter, $b \approx 6 R_\odot$ (see both \cite{Bertotti:2003rm} and \cite{Ashby:2009bb}) and the measured value $\gamma_\text{\tiny PPN}=1+(2.1 \pm 2.3)\times10^{-5}$, we obtain a naive constraint of $\L \gtrsim {\cal O}(10^{-19}) \text{MeV}$, which is 14 orders of magnitude better than the above naive cosmological bound on $\L_2$ alone. However, 
just like in the cosmological case, we need to recognize that the energy of the radio waves used in the Cassini observations is roughly 10 GHz $\approx 4 \times 10^{-11}$ MeV, at least 8 orders of magnitude greater than the lower bound. Once again, we need to ask instead how accurate the timing measurement needs to be to probe $\Lambda_{1,2} \gtrsim 10$ GHz.  Using the second term on the right hand side of \eqref{Modified Delay}, the answer is $\Delta t \lesssim \mathcal{O}(10^{-27})$s. This is at least 18 orders of magnitude more precise than the Casinni observation, if one estimates the fractional error for the latter observation to be given by the current bound on $\gamma_\text{\tiny PPN}$.\textsuperscript{\footnotemark[3]}\footnotetext[3]{One may consider using the timing measurements of pulsars, with masses of order $M_{\odot}$ and radii on the order of several km, as was exploited by the authors of \cite{Prasanna:2003ix}.  However, there the measured Shapiro delay (really $r$, the range of Shapiro delay) is governed not by the radius of the pulsar itself, as they have claimed, but rather the separation distance between pulsar and companion (see \cite{Backer:1986wa} and \cite{Stairs:2003eg}), which is typically of order $R_{\odot}$. Considering the relative errors on such observations, solar system observations remain a superior test.}

\section{Summary and Discussion}\label{sec:V}
We have constructed the most general effective Lagrangian coupling electromagnetism and gravity up to mass dimension 6, built from all possible contractions between tensors that obey the underlying gauge symmetries of both theories. There are many such non-minimal terms.  However, after allowing for field re-definitions of the electromagnetic vector potential and the metric, it is seen that the number of non-redundant terms reduces to two.  One represents the type of coupling already explored from one-loop quantum effects in QED.  The other is parity-violating; if it is induced by the standard model, we expect it to come from the electroweak sector and to be suppressed by $\mathcal{O}(M_\text{W}^{-2})$.

We have also discussed some of the phenomenology of these non-minimal terms, including birefringence of the CMB, modified dispersion relations for the photon, as well as corrections to the Shapiro time delay, and deflection angle. Via detailed calculations, we came to see that cosmological and solar system probes do not seem likely, within the foreseeable future, to give any physically useful constraints on $1/\L_1$ and $1/\L_2$. This is because, as already alluded to in the introduction, observations are 
unlikely to ever reach a level of precision to even probe $\L_{1,2}$-scales of the same magnitude of the photon energies involved.

We end with some suggestions on possible future work.  Other than the weak lensing surveys already mentioned in the body of the paper, Drummond and Hathrell \cite{Drummond:1979pp} have initiated the investigation of these modified photon dynamics on a gravitational wave background; it would be natural to extend their analysis to include the effects of the parity violating $1/\L_2^{\phantom{2}2}$ term. In this paper, we have only examined the dynamics of the photon itself; looking at how Einstein's field equations are altered and their corresponding implications may provide alternate channels to constrain $\L_1$ and $\L_2$.  One may also want to seek perturbative solutions for $A_\mu$ or $F_{\mu\nu}$ (with $1/\L_{1,2}^{\phantom{1,2}2} \neq 0$) about exact solutions of the Einstein-Maxwell system (with $1/\L_{1,2}^{\phantom{1,2}2}=0$) containing pure magnetic fields, as a toy model of more realistic astrophysical systems. Finally, a stability analysis of the full system in \eqref{FinalAction} may also be performed to perhaps help constrain the range of physically reasonable values of $1/\L_{1,2}^{\phantom{1,2}2}$.

\begin{acknowledgments}
We would like to thank Simeon Hellerman and Lawrence Widrow for discussions; as well as Luciano Iess, Bruno Bertotti, and Neil Ashby for their assistance in understanding issues related to the Shapiro delay. We also would like to acknowledge Alex Vikman for raising the issue of stability.
\end{acknowledgments}

\appendix
\sec{Shapiro delay calculations}\label{app}
For the idealized case where the Earth, satellite and Sun are all motionless, the proper Earth-satellite time of flight for a light signal (as measured on the Earth) has been first computed by Shapiro \cite{Shapiro:1964uw} and elaborated in detail in the standard textbook by Weinberg \cite{weinberg} in the standard Schwarzschild gauge. Subsequent calculations in isotropic (spatially-conformally-flat) coordinates have also been done (see \cite{MTW:1973},\cite{Will:2005va},\cite{Bertotti:2003rm}).  At order $r_s$ there appear to be differences amongst these calculations, but these can be attributed to either the choice of gauge or whether or not the straight line approximation is used.  Reference \cite{Ashby:2009bb} offers a nice discussion and interpolation between some of the methods.

Most of the calculations found in the literature mentioned above, except in \cite{weinberg}, use the straight line approximation to compute the time delay.  This method is, in fact, exact up to order $r_s$, at least in the idealized case of motionless bodies, because of Fermat's principle in a static spacetime ($\partial_t g_{\mu\nu} = g_{0i} = 0$).  Namely, the coordinate time of flight
\begin{equation}
\Delta t = \int \sqrt{-\f{g_{ij}}{g_{00}}\f{d x^i}{ds}\f{d x^j}{ds}} d s
\end{equation}
is extremized if $x^\mu(s)$ is a null geodesic of the spacetime described by $g_{\m\n}$.  In a weakly curved spacetime where $r_s$ is much smaller than all other length scales, both the null geodesics and $\Delta t$ can be developed as a power series in $r_s$. The $\O{r_s}$ accurate $\Delta t$ can be obtained by employing the lowest order solution to the null geodesic equation, which is simply a straight line.  The contribution to $\Delta t$ due to the deviation of the null path from a straight line begins at $\O{r_s^2}$, due to Fermat's principle.

One could show the equivalence between Weinberg's \cite{weinberg} and Shapiro's \cite{Shapiro:1964uw} formulas by an explicit calculation, in which the true distance of closest approach, $r_0$, and the ``straight line" distance of closest approach, $b$, are related through the light deflection angle integral \eqref{deltaphi}.

\bibliographystyle{h-physrev}
\bibliography{EM-grav}
\end{document}